\documentstyle[12pt,axodraw]{article}
\newcommand{\be}{\begin{equation}}
\newcommand{\bea}{\begin{eqnarray}}
\newcommand{\eea}{\end{eqnarray}}
\newcommand{\ba}{\begin{array}}
\newcommand{\ea}{\end{array}}
\newcommand{\ee}{\end{equation}}

\expandafter\ifx\csname mathbbm\endcsname\relax

\else

\fi
\begin{document}
\begin{titlepage}
\hfill
\vbox{
    \halign{#\hfil         \cr
           IPM/P-2003/054 \cr
           hep-th/0309037  \cr
           } 
      }  
\vspace*{20mm}
\begin{center}
{\Large {\bf One-Loop Perturbative Corrections to 
non(anti)commutativity Parameter of ${\cal N}={1\over 2}$ Supersymmetric 
$U(N)$ Gauge Theory\\ }}

\vspace*{15mm}
\vspace*{1mm}
{Mohsen Alishahiha$^a$ \footnote{alishah@theory.ipm.ac.ir},
Ahmad Ghodsi$^a$ \footnote{ahmad@theory.ipm.ac.ir}
and N\'eda Sadooghi$^{a,b}$ \footnote{sadooghi@theory.ipm.ac.ir}
} \\
\vspace*{1cm}

{\it$^a$ Institute for Studies in Theoretical Physics
and Mathematics (IPM)\\
P.O. Box 19395-5531, Tehran, Iran \\ \vspace{3mm}
$^b$ Department of Physics, Sharif University of Technology\\
P.O. Box 11365-9161, Tehran, Iran}\\

\vspace*{1cm}
\end{center}

\begin{abstract}
Perturbative corrections to ${\cal N}={1\over 2}$
supersymmetric $U(N)$ gauge theory at one-loop order are studied. It is  shown 
that whereas the 
quantum corrections 
to ${\cal N}=1$ sector of the theory are not affected by the $C$-deformation, 
the non(anti)commutativity parameter $C$ receives one-loop perturbative 
corrections. These perturbative corrections are computed by performing an 
explicit one-loop 
calculation of the three and 
four-point functions of the theory. The running of the 
non\-(anti)\-com\-mu\-ta\-ti\-vi\-ty parameter $C$ is also studied using an 
appropriate Callan-Symanzik equation.
\end{abstract}
\vspace{1cm}
\end{titlepage}

\section{Introduction}
In the past few years noncommutative field theories have been studied 
extensively, 
mainly due to their realization in the string theory  
\cite{Connes:1997cr}. To be more  precise, by wrapping the branes with non-zero 
constant background $B_{\mu\nu}$ field the
corresponding low energy effective gauge theory is deformed to a
noncommutative supersymmetric gauge theory in such a way that 
those (bosonic) directions in which the $B_{\mu\nu}$ field is defined become
noncommutative. The noncommutativity parameter can then be given in
terms of the finite $B_{\mu\nu}$ background field
 \cite{Seiberg:1999vs}-\cite{Schomerus:1999ug}.
\par
Recently it has been shown that noncommutative superspace is also realized in 
the string theory by turning on a constant graviphoton field strength 
$F^{\alpha\beta}$
\cite{{Ooguri:2003qp},{Ooguri:2003tt},{Seiberg:2003yz}}, which now changes the
anticommutation relation between Grassmanian (fermionic) variables of the
superspace. This deformation is in such a way that the anticommuting 
coordinates $\theta$
form a Clifford algebra  
\cite{Casalbuoni:1975bj}-\cite{Alday:2003ms}
\be 
\{\theta^\alpha ,
\theta^\beta\}=2{\alpha'}^2F^{\alpha\beta}=C^{\alpha\beta}. 
\label{non}
\ee
Starting from an ${\cal{N}}=1$ supersymmetric gauge theory, the half of the 
supersymmetry is therefore broken by this deformation and  we are 
left with an 
${\cal{N}}=1/2$ supersymmetric gauge theory \cite{Seiberg:2003yz}.  Note 
that since the anticommutation relation of ${\bar \theta}$ remains
undeformed, $\bar{\theta}$ is not the complex conjugate of $\theta$ and this 
is only possible in the  Euclidean space ${\rm R}^4$, where the ordinary 
space-time coordinates $x_{\mu}$ turn out to be noncommutative. 
In fact one has\footnote{The noncommutativity of the $x$ space due to the 
RR fields is also studied in \cite{Cornalba:2002cu}.}  
\be
[x^\mu,\theta^\alpha]=iC^{\alpha\beta}\sigma^\mu_{\beta{\dot
\alpha}} {\bar \theta}^{{\dot
\alpha}},\;\;\;\;\;\;[x^\mu,x^\nu]={\bar \theta} {\bar
\theta}C^{\mu\nu}\;, 
\ee 
where
$C^{\mu\nu}=C^{\alpha\beta}\epsilon_{\beta\gamma}
\sigma_{\alpha}^{\mu\nu\;\gamma}$. The chiral coordinates
$y^\mu=x^\mu+i\theta^\alpha
\sigma^\mu_{\alpha {\dot \alpha}}
{\bar \theta}^{{\dot \alpha}}$, however, can be taken to be commutative
\be 
[y^\mu,y^\nu]=[y^\mu,\theta^\alpha]=[y^\mu,{\bar
\theta}^{{\dot \alpha}}]=0\;. 
\ee  
One of the consequences of the non(anti)commutation relation (\ref{non}) in
the superspace is that the products of superfields as functions of
$\theta$ are now to be ordered. This can be imposed by a novel $*$-product 
defined by 
\be
f(\theta)*g(\theta)=f(\theta)e^{-{C^{\alpha\beta}\over 2}
\stackrel{\leftarrow}{\partial}_\alpha
\stackrel{\rightarrow}{\partial}_\beta}g(\theta)\;.
\ee
Replacing all the ordinary products with the above $*$-product, one may proceed 
by studying a supersymmetric field theory in this 
non(anti)commuting superspace, taking into account that this deformed 
supersymmetry
algebra admits  well-defined representations. Namely, one can
define chiral and vector superfields much similar to the ordinary
${\cal N}=1$ supersymmetry \cite{Seiberg:2003yz}. For example, 
the vector multiplet in Wess-Zumino gauge is given by 
\cite{Seiberg:2003yz}
\newpage
\bea
V(y,\theta,{\bar \theta})&=&-\theta\sigma^\mu{\bar\theta}A_\mu
+i\theta\theta{\bar\theta}{\bar \lambda}-i{\bar\theta}
{\bar\theta}\theta^\alpha\left(\lambda_\alpha+{1\over 4}
\epsilon_{\alpha\beta}C^{\beta\gamma}\sigma^\mu_{\gamma{\dot\gamma}}
\{{\bar \lambda}^{\dot \gamma},A_{\mu}\}\right)\cr &&\cr
&&+{1\over 2}{\theta}{\theta}{\bar\theta}{\bar\theta}
\left(D-i\partial_\mu A^\mu\right)\;.
\eea
One can also define the corresponding superfield strength tensor $W_{\alpha}$ 
and thereby give 
\bea
\int d^2\theta\;{\rm Tr}\  W*W&=&\int d^2\theta\;{\rm Tr}\ WW(C=0)
-iC^{\mu\nu}{\rm Tr}\ (F_{\mu\nu}{\bar \lambda}{\bar \lambda})
+{|C|^2\over 4}{\rm Tr}\ ({\bar \lambda}{\bar \lambda})^2\cr &&\cr
\int d^2{\bar \theta}\;{\rm Tr}\  {\bar W}*{\bar W}&=&
\int d^2\theta\;{\rm Tr}\ {\bar W}{\bar W}(C=0)
-iC^{\mu\nu}{\rm Tr}\ (F_{\mu\nu}{\bar \lambda}{\bar \lambda})
+{|C|^2\over 4}{\rm Tr}\ ({\bar \lambda}{\bar \lambda})^2\cr &&\cr
&&+{\rm total\;derivative}\;,
\eea
which can be used to define the Lagrangian of ${\cal{N}}=1/2$ supersymmetric 
U(N) gauge theory. 
\par
Recently various field theoretical aspects of the non(anti)commuting  
superspace
have been studied in \cite{Britto:2003aj1}-
\cite{Imaanpur:2003jj}. In particular, the renormalizability of 
non\-(anti)\-com\-mu\-ta\-tive gauge theory with ${\cal N}=1/2$ supersymmetry 
has been studied in \cite{Lunin:2003bm}.  Using an explicit dimensional 
analysis, the authors in 
\cite{Lunin:2003bm} show that this theory is renormalizable to all
order of perturbation theory. 
\par
In this paper, an explicit one-loop perturbative calculation of ${\cal{N}}=1/2$ 
supersymmetric $U(N)$ gauge theory is performed. It is shown that whereas the 
quantum corrections to ${\cal{N}}=1$ sector remain unaffected by the 
$C$-deformation, the non-(anti)commutativity parameter $C$ itself receives 
one-loop quantum corrections. An explicit one-loop calculation of the 
three and four-point functions of the theory in the $C$-deformed sector is 
carried out to calculate these corrections explicitly. 
In Section 2, after 
giving the full action of the theory including the pure gauge part, the 
gauge fixing and the ghost parts, the Feynman rules for propagators and 
vertices are presented. Using these Feynman rules an explicit 
one-loop calculation of the undeformed ${\cal{N}}=1$ sector and the 
$C$-deformed sector 
of the 
theory is performed in Section 3. 
We have shown that the 
corresponding renormalization constant $Z_{C}$ is {\it always}
given by the inverse renormalization constant $Z_{g}$ which 
corresponds to the coupling constant $g$ 
of the theory. To find the one-loop correction to $C^{2}$ and to check the 
relation $Z_{C^{2}}=Z_{C}^{2}$, suggested 
first in \cite{Lunin:2003bm}, an explicit one-loop calculation 
of the $\bar{\lambda}$ four-point function of the theory is performed in 
Section 4.       
In Section 5, we derive an appropriate  Callan-Symanzik differential 
equation for the renormalized three-point function 
$\Gamma_{A\bar{\lambda}\bar{\lambda}}$. 
The running of the non(anti)commutativity parameter 
$C$ with the RG-scale $\mu$ is then calculated explicitly by 
solving the corresponding  Callan-Symanzik $\gamma_{C}$-function for $C$. We 
have shown that the 
product of $C(\mu)g(\mu)$ is an RG-invariant. Here $g(\mu)$ is the standard 
running coupling constant of the ordinary $U(N)$ theory.
The last section is devoted to conclusion.
\par
The following conventions are used in this paper. For $U(N)$ 
elements we use capital
indices while for $SU(N)$ elements small indices are used so
that
\be
{\rm Tr}(t^A t^B)={1\over2}\delta^{AB},\;\;\;\;
[t^A,t^B]=if^{ABC}t^C,\;\;\;\;\{t^A,t^B\}=d^{ABC}t^C\;.
\ee
In this notation we have $t^0={1\over \sqrt{2N}},\;
d^{0AB}=\sqrt{{2\over N}}\delta^{AB}\;$. We further use
\cite{{Armoni:2000xr},{Bonora:2000ga}} 
\be
f^{IAJ}f^{JBK}f^{KCI}=-{N\over 2}f^{ABC}, \;\;\;\;
d^{IAJ}f^{JBK}f^{KCI}=-{N\over 2}d^{ABC}d_Af_Bf_C\;,
\label{convension}
\ee
and 
\bea
&&f^{IAJ}f^{JBK}d^{KCL}d^{LDI}={1\over 2}f_Af_B\bigg{[}f_Cf_D\left(
\delta^{AC}\delta^{BD}-\delta^{AB}\delta^{CD}+
\delta^{AD}\delta^{BC}\right)\cr &&\cr 
&&+{N\over 4}d_Cd_D\left(f^{ABX}f^{CDX}-
f^{ADX}f^{BCX}-d^{ABX}d^{CDX}-d^{ADX}d^{BCX}\right)\bigg{]}
\eea

where $f_A=1-\delta_{0A},\;d_A=2-f_{A}$.
Note also that $f^{iaj}f^{jbi} =-N\delta^{ab}$.
\section{${\cal N}={1\over 2},\ U(N)$ SYM theory; Feynman Rules}
Following our notation from the previous section, the classical action
for ${\cal N}={1\over 2}$ supersymmetric $U(N)$ gauge theory is given by
\footnote{In comparison with \cite{Seiberg:2003yz} we have rescaled
the components of $V$ by $-2g$.}
\be
{\cal{S}}_{\rm gauge}=\int d^4x\;{\rm Tr}\ \bigg[-{1\over 
2}F^{\mu\nu}F_{\mu\nu}-2i{\bar 
\lambda}
{\bar \sigma}^\mu {\cal D}_{\mu}\lambda +D^2+2igC^{\mu\nu}
F_{\mu\nu}{\bar \lambda}{\bar \lambda}
+g^2|C|^2({\bar \lambda}{\bar \lambda})^2
\bigg]\;,
\label{AC}
\ee
with 
\bea
F^A_{\mu\nu}&=&\partial_{\mu}A^A_{\nu}-\partial_{\nu}A^A_{\mu}+gf^{ABC}A^B_{\mu}
A^C_{\nu}\;,\cr &&\cr
{\cal D}_{\mu}\lambda^A&=&\partial_{\mu}\lambda^A+gf^{ABC}A_\mu^B\lambda^C\;.
\eea
In the superspace formalism the gauge fixing action is given by
\be
{\cal{S}}_{\rm GF}=-{1\over 4\xi}\int d^4\theta\;{\rm Tr}(D^2V*{\bar D}^2V)\;,
\ee
that in terms of the components of the vector superfield reads
\be
{\cal{S}}_{\rm GF}=-{1\over \xi}
{\rm Tr}\left(D^2+(\partial_\mu A^\mu)^2
-2i{\bar \lambda}{\bar \sigma}^\mu \partial_\mu\lambda+{i\over 2}gC^{\mu\nu}
\partial_{[\mu}A_{\nu]}({\bar \lambda}{\bar\lambda})\right).
\label{GFAC}
\ee
Together with the  gauge invariant action (\ref{AC}), we have
\bea
\lefteqn{
{\cal{S}}_{gauge}+{\cal{S}}_{\rm GF}=}\nonumber\\
&=&\int d^4x\;{\rm 
Tr}\bigg[-{1\over 
2}F^{\mu\nu}F_{\mu\nu}-{1\over \xi}
(\partial_\mu A^\mu)^2+(1-{1\over \xi})D^2
-2i{\bar \lambda}
{\bar \sigma}^\mu ({\cal D}_{\mu}-{1\over \xi}\partial_\mu)\lambda \cr &&\cr
&&+2igC^{\mu\nu}
(F_{\mu\nu}-{1\over 4\xi}\partial_{[\mu} A_{\nu]}){\bar \lambda}{\bar \lambda}
+g^2|C|^2({\bar \lambda}{\bar \lambda})^2
\bigg]\;.
\eea
This action can be used to read the Feynman rules. In the super-Fermi-Feynman 
gauge, $\xi=1$,  the Feynman rules are given by
\begin{eqnarray}
\SetScale{0.8}
    \begin{picture}(80,20)(0,0)
    \Vertex(-5,0){2}
    \ArrowLine(-5,0)(45,0)
    \Vertex(45,0){2}
    \Text(20,10)[]{$p$}
    \Text(-5,-10)[]{${\bar \lambda}_{{\dot \alpha}}^A$}
    \Text(40,-10)[]{$\lambda_\alpha^B$}
    \end{picture}
\hspace{5.4cm} \;\;\;\;\;\;\;\;\;
-\frac{\delta^{AB}\sigma^{\mu}_{\alpha\dot{\alpha}}p_{\mu}}{p^{2}},
\end{eqnarray}
\vspace{0.3cm}
\begin{eqnarray}\label{FX1}
\SetScale{0.8}
    \begin{picture}(80,20)(0,0)
    \Vertex(-20,0){2}
    \Gluon(-20,0)(30,0){3}{6}
    \Vertex(30,0){2}
    \Text(0,10)[]{$k$}
    \Text(-20,-10)[]{$A_\mu^A$}
    \Text(25,-10)[]{$A_\nu^B$}
    \end{picture}
\hspace{5cm} \;\;\;\;\;\;\;\;\;\;\;-\frac{\eta_{\mu\nu}\delta^{AB}}{k^{2}},
\end{eqnarray}
\vspace{0.3cm}
\begin{eqnarray}\label{FX3}
\SetScale{0.8}
    \begin{picture}(80,20)(0,0)
        \Vertex(0,0){2}
    \ArrowArc(0,-30)(15,40,140)
    \Gluon(0,0)(0,20){2}{4}
    \LongArrow(5,18)(5,12)
    \ArrowLine(0,0)(-20,-20)
    \ArrowLine(20,-20)(0,0)
    \Text(20,15)[]{$A^{B}_{\mu}$}
    \Text(-25,-15)[]{$\lambda^{ {\alpha}C}$}
    \Text(25,-15)[]{$\bar{\lambda}^{ {\dot{\alpha}}A}$}
    \end{picture}
\hspace{5.6cm}\;\;\; ig\ f^{ABC}\bar{\sigma}^{\mu\ \dot{\alpha}\alpha}\;,
\end{eqnarray}
\vspace{0.3cm}
\begin{eqnarray}\label{FX4}
\SetScale{0.8}
    \begin{picture}(80,20)(0,0)
        \Vertex(0,0){2}
    \Gluon(0,0)(0,20){2}{4}
    \LongArrow(5,18)(5,12)
    \ArrowLine(0,0)(-20,-20)
    \ArrowLine(0,0)(20,-20)
    \ArrowArc(0,-30)(15,40,140)
    \Text(20,15)[]{$A_{\nu}^{C}, k$}
    \Text(-25,-15)[]{$\bar{\lambda}^{B {\dot{\beta}}}$}
    \Text(25,-15)[]{$\bar{\lambda}^{A {\dot{\alpha}}}$}
    \end{picture}
\hspace{4.8cm}
-{3\over 2}gC^{\mu\nu}k_{\mu}\epsilon^{\dot{\alpha}\dot{\beta}}d^{ABC}\;,
\end{eqnarray}
\vspace{0.3cm}
\begin{eqnarray}\label{FX1d}
\SetScale{0.8}
    \begin{picture}(50,0)(0,0)
    \Vertex(0,0){2}
    \Gluon(0,0)(0,20){2}{4}
    \LongArrow(-10,20)(-10,12)
    \LongArrow(-25,-15)(-20,-8)
    \LongArrow(25,-15)(20,-8)
    \Gluon(-20,-20)(0,0){2}{4}
    \Gluon(20,-20)(0,0){2}{4}
    \Text(25,15)[]{$A_\mu^A,k$}
    \Text(-40,-15)[]{$A_\nu^B,p$}
    \Text(40,-15)[]{$A_\rho^C,q$}
    \end{picture}
\hspace{0cm}\;\;\; igf^{ABC}\bigg{(}\eta^{\mu\nu}(k-p)^\rho+\eta^{\nu\rho}
(p-q)^\mu
+\eta^{\rho\mu}(q-k)^\nu\;\bigg{)}\;,
\end{eqnarray}
\vspace{0.3cm}
\begin{eqnarray}
\SetScale{0.8}
    \begin{picture}(50,0)(0,0)
    \Vertex(0,0){2}
    \ArrowLine(0,0)(20,20)
    \ArrowLine(0,0)(-20,20)
    \Gluon(-20,-20)(0,0){2}{4}
    \Gluon(20,-20)(0,0){2}{4}
    \ArrowArc(0,30)(15,230,-50)
    \LongArrow(-25,-15)(-20,-8)
    \LongArrow(25,-15)(20,-8)
    \Text(-30,15)[]{$\bar{\lambda}^{ {\dot{\alpha}}A}$}
    \Text(30,15)[]{$\bar{\lambda}^{ {\dot{\beta}}B}$}
    \Text(-30,-15)[]{$A_{\mu}^{C}$}
    \Text(30,-15)[]{$A_{\nu}^D$}
    \end{picture}
\hspace{5cm} 2ig^2C^{\mu\nu}\epsilon^{{\dot \alpha}{\dot \beta}}
d^{ABL}f^{LCD},
\end{eqnarray}
\vspace{0.3cm}
\begin{eqnarray}
\label{FX3d}
\SetScale{0.8}
    \begin{picture}(50,0)(0,0)
    \Vertex(0,0){2}
    \ArrowLine(0,0)(20,20)
    \ArrowLine(0,0)(-20,20)
    \ArrowLine(0,0)(-20,-20)
    \ArrowLine(0,0)(20,-20)
    \Text(-30,15)[]{$\bar{\lambda}^{\dot{\alpha}A}$}
    \Text(30,15)[]{$\bar{\lambda}^{\dot{\delta}D}$}
    \Text(-30,-15)[]{$\bar{\lambda}^{\dot{\beta}B}$}
    \Text(30,-15)[]{$\bar{\lambda}^{\dot{\gamma}C}$}
    \end{picture}
\hspace{.5cm} g^2|C|^2\bigg{(}\;\epsilon^{{\dot \alpha}{\dot \beta}}
\epsilon^{{\dot \gamma}{\dot \delta}}d^{ABM}d^{MCD}
    -\epsilon^{{\dot \alpha}{\dot \gamma}}
\epsilon^{{\dot \beta}{\dot \delta}}d^{ACM}d^{MBD}\cr
     +\epsilon^{{\dot \alpha}{\dot \delta}}
\epsilon^{{\dot \beta}{\dot \gamma}}d^{ADM}d^{MBC}\;\bigg{)}\;.
\end{eqnarray}
Further, the ghost field 
action in the superfield formalism is given by
\be
{\cal{S}}_{\rm ghost}=2\ {\rm Tr}\ \int d^4\theta\;({\mathcal{C}}'+{\bar 
{\mathcal{C}}}')L_{-gV}\left[({\mathcal{C}}+{\bar {\mathcal{C}}})+\coth(L_{-gV})
\;({\mathcal{C}}-{\bar {\mathcal{C}}})\right]\;,
\ee
where $L_XY=[X,Y]$ and all products are understood as star-product.
We use the following notations for the components of the ghost superfields 
\bea
&&{\mathcal{C}}=c+\sqrt{2}\theta\zeta+\theta\theta F,\hspace{1cm}
{\mathcal{C}}'=b+\sqrt{2}\theta\eta+\theta\theta F'\;\cr &&\cr
&&\bar{{\mathcal{C}}}=\bar{c}+\sqrt{2}\bar{\theta}
\bar{\zeta}-2i\theta\sigma^{\mu}
\bar{\theta}\partial_{\mu}\bar{c}+\overline{\theta\theta} (\bar{F}+
i\sqrt{2}\theta\sigma^{\mu}
\partial_{\mu}\bar{\zeta}+\theta\theta\partial^2\bar{c}),\cr
&&\cr
&&\bar{{\mathcal{C}}}'=\bar{b}+\sqrt{2}\bar{\theta}
\bar{\eta}-2i\theta\sigma^{\mu}\bar{\theta}\partial_{\mu}\bar{b}
+\overline{\theta\theta} (\bar{F}'+i\sqrt{2}\theta\sigma^{\mu}
\partial_{\mu}\bar{\eta}
+\theta\theta\partial^2\bar{b}).
\eea
After integrating out the auxiliary field, the quadratic (kinetic)
term of the ghost action reads
\begin{eqnarray}\label{kinetic}
2\ {\rm Tr}\ \int d^4\theta\;({\bar {\mathcal{C}}}'{\mathcal{C}}-
{\mathcal{C}}'{\bar {\mathcal{C}}})=
-2\ {\rm Tr}\ (c\partial^2{\bar b}+b\partial^2
{\bar c}+i\zeta\sigma^\mu\partial_\mu{\bar \eta}+
i\eta\sigma^\mu\partial_\mu{\bar \zeta})\;.
\end{eqnarray}
The interaction terms of the ghost action have two different parts.
The first one arises from those terms including one vector superfield
\bea\label{int1}
&&-2g\ {\rm Tr}\ \int d^4\theta({\mathcal{C}}'
+{\bar {\mathcal{C}}}')[V,{\mathcal{C}}+{\bar {\mathcal{C}}}]=
gf^{ABC}\bigg[\left(\partial^\mu
{\bar b}^Ac^B+\partial^\mu{\bar c}^Ab^B\right)A_{\mu}^C-\bar{b}^A
\bar{c}^B\partial^{\mu}\!
A^C_\mu\bigg ]
\cr &&\cr &&+\left({i\over 2}
\zeta^A\sigma^\mu {\bar \eta}^B+{i\over 2}\eta^A \sigma^\mu
{\bar \zeta}^B\right)A_{\mu}^C+{\sqrt{2}\over 2}gf^{ABC}
\left((b+{\bar b})^A\zeta^{\alpha B}
+(c+{\bar c})^A\eta^{\alpha B}\right)f^C_{\alpha}\cr &&\cr
&&-{\sqrt{2}\over 2}gf^{ABC}\left((b+{\bar b})^A
{\bar \zeta}_{{\dot \alpha}}^B
+(c+{\bar c})^A{\bar \eta}_{{\dot \alpha}}^B\right)
{\bar \lambda}^{{\dot \alpha} C}
\cr&&\cr &&+
{\sqrt{2}\over 2}gd^{ABC}\sigma^\mu_{\alpha{\dot \alpha}}
C^{\alpha\beta}
\left(\partial_\mu{\bar b}^A\zeta^B_{\beta}+\partial_\mu
{\bar c}^A
\eta^B_\beta\right){\bar \lambda}^{{\dot \alpha} C}\;,
\eea
where $f^C_\alpha=\lambda^C_{\alpha}-{g\over 2}
\epsilon_{\alpha\beta}C^{\beta\gamma}
\sigma^{\mu}_{\gamma{\dot \gamma}}d^{CDE}{\bar \lambda}^{{\dot \gamma}D}
A^E_{\mu}$.
Further, for our purpose, the relevant terms of the second part of the ghost 
action, including two vector superfields read
\bea\label{int2}
&&\frac{2}{3}g^2\ {\rm Tr}\ \int  d^4\theta ({\mathcal{C}}'+
{\bar {\mathcal{C}}}')
[\;V,\;[V,{\mathcal{C}}-{\bar {\mathcal{C}}}]\;]=
{2\over3}g^{2}\ {\rm Tr}\ \bigg{[}
-i{\sqrt{2}\over4}(b+\bar{b})\{F_{1\alpha},\zeta^\alpha\}\cr &&\cr
&&\;\;\;\;\;\;\;\;\;\;\;\;\;
+i{\sqrt{2}\over4}\eta_\alpha [F_1^\alpha ,(c-\bar{c})]
-\eta_\alpha \{B_1,\zeta^\alpha\}
+{|C|^2\over4}\eta_\alpha\bar{\lambda}_{\dot{\alpha}}
\zeta^\alpha\bar{\lambda}^{\dot{\alpha}}\cr &&\cr
&&\;\;\;\;\;\;\;\;\;\;\;\;\;
+{i\sqrt{2}\over2}C^{\alpha\beta}\sigma_{\alpha\dot{\alpha}}^\mu
\bigg((b+\bar{b})(A_\mu\zeta_\beta
\bar{\lambda}^{\dot{\alpha}}-\bar{\lambda}^{\dot{\alpha}}
\zeta_\beta A_\mu)\cr &&\cr
&&\;\;\;\;\;\;\;\;\;\;\;\;\;
+\eta_\beta A_\mu (c-\bar{c})\bar{\lambda}^{\dot{\alpha}}+
\eta_\beta\bar{\lambda}^{\dot{\alpha}}(c-\bar{c})A_\mu
\bigg)+\cdots\bigg{]},
\eea
with
\be
B_1=-{1\over8}|C|^2\overline{\lambda\lambda}, 
\hspace{1cm}\mbox{and}\hspace{1cm}
F^\alpha_1=C^{\alpha\beta}\sigma_{\beta\dot{\alpha}}^\mu[A_\mu ,
\bar{\lambda}^{\dot{\alpha}}].
\ee
Using the kinetic part of the ghost action (\ref{kinetic}) the propagators 
read
\begin{eqnarray}
\SetScale{0.8}
    \begin{picture}(80,20)(0,0)
    \Vertex(-20,0){2}
    \DashArrowLine(-20,0)(30,0){4}
    \Vertex(30,0){2}
    \Text(0,10)[]{$k$}
    \Text(-20,-10)[]{$b^A,c^A$}
    \Text(25,-10)[]{${\bar c}^B,{\bar b}^B$}
    \end{picture}
\hspace{5cm} \;\;\;\;\;\;\;\;\;\;\;-\frac{\delta^{AB}}{k^{2}},
\end{eqnarray}
\begin{eqnarray}
\SetScale{0.8}
    \begin{picture}(80,20)(0,0)
    \Vertex(-20,0){2}
    \DashArrowLine(-20,0)(30,0){2}
    \Vertex(30,0){2}
    \Text(0,10)[]{$k$}
    \Text(25,-10)[]{${\bar \zeta}^B_{\dot \alpha},{\bar \eta}^B_{\dot \alpha}$}
    \Text(-20,-10)[]{$\eta^A_\alpha,\zeta^A_\alpha$}
    \end{picture}
\hspace{5cm} -\frac{\delta^{AB}
{\bar \sigma}^{\mu\;{\dot\alpha}\alpha}k_\mu}{k^{2}},
\end{eqnarray}
\vspace{0.4cm}
The relevant vertices can be read from the interaction parts of the 
action  (\ref{int1}) and (\ref{int2}) 
\begin{eqnarray}
\SetScale{0.8}
    \begin{picture}(80,20)(0,0)
        \Vertex(0,0){2}
    \Gluon(0,0)(0,20){2}{4}
    \LongArrow(5,18)(5,12)
    \DashLine(0,0)(-20,-20){4}
    \DashLine(0,0)(20,-20){4}
    \ArrowArc(0,-30)(15,40,140)
    \Text(20,15)[]{$A_{\mu}^{B}$}
    \Text(-25,-15)[]{$k,{\bar c}^C$}
    \Text(25,-15)[]{$b^A$}
    \end{picture}
\hspace{4.8cm}
\;\;\;\;\;\;\;\;igf^{ABC}k_\mu,
\end{eqnarray}
\vspace{0.3cm}

\begin{eqnarray}
\SetScale{0.8}
    \begin{picture}(80,20)(0,0)
        \Vertex(0,0){2}
    \Gluon(0,0)(0,20){2}{4}
    \LongArrow(5,18)(5,12)
    \DashLine(0,0)(-20,-20){4}
    \DashLine(0,0)(20,-20){4}
    \ArrowArc(0,-30)(15,40,140)
    \Text(20,15)[]{$A_{\mu}^B$}
    \Text(-25,-15)[]{$c^C$}
    \Text(28,-15)[]{${\bar b}^A,k$}
    \end{picture}
\hspace{4.8cm}
\;\;\;\;\;\;\;\;igf^{ABC}k_\mu,
\end{eqnarray}
\vspace{0.3cm}
\begin{eqnarray}
\SetScale{0.8}
    \begin{picture}(80,20)(0,0)
        \Vertex(0,0){2}
    \Gluon(0,0)(0,20){2}{4}
    \LongArrow(5,18)(5,12)
    \DashLine(0,0)(-20,-20){2}
    \DashLine(0,0)(20,-20){2}
    \Text(20,15)[]{$A_{\mu}^{B}$}
    \Text(-25,-15)[]{$\zeta^C_\alpha$}
    \Text(25,-15)[]{${\bar \eta}^A_{{\dot \alpha}}$}
    \end{picture}
\hspace{4.8cm}
-{i\over 2}gf^{ABC}\sigma^{\mu}_{\alpha{\dot \alpha}},
\end{eqnarray}
\vspace{0.3cm}
\begin{eqnarray}
\SetScale{0.8}
    \begin{picture}(80,20)(0,0)
        \Vertex(0,0){2}
    \Gluon(0,0)(0,20){2}{4}
    \LongArrow(5,18)(5,12)
    \DashLine(0,0)(-20,-20){2}
    \DashLine(0,0)(20,-20){2}
    \Text(20,15)[]{$A_{\mu}^{B}$}
    \Text(-25,-15)[]{${\bar \zeta}^C_{{\dot \alpha}}$}
    \Text(25,-15)[]{$\eta^A_\alpha$}
    \end{picture}
\hspace{5.2cm}
{i\over 2}gf^{ABC}\sigma^{\mu}_{\alpha{\dot \alpha}},
\end{eqnarray}
\vspace{0.3cm}
\begin{eqnarray}
\SetScale{0.8}
    \begin{picture}(80,20)(0,0)
        \Vertex(0,0){2}
    \ArrowLine(0,20)(0,0)
    \DashLine(0,0)(-20,-20){2}
    \DashLine(0,0)(20,-20){4}
    \ArrowArc(0,-30)(15,40,140)
    \Text(20,15)[]{$\lambda^B_\beta$}
    \Text(-25,-15)[]{$\Psi^C _\alpha$}
    \Text(25,-15)[]{$A^A$}
    \end{picture}
\hspace{4.5cm}
-{\sqrt{2}\over 2}gf^{ABC}\epsilon_{\alpha\beta},
\end{eqnarray}
\vspace{0.3cm}
\begin{eqnarray}
\SetScale{0.8}
    \begin{picture}(80,20)(0,0)
        \Vertex(0,0){2}
    \ArrowLine(0,0)(0,20)
    \DashLine(0,0)(-20,-20){2}
    \DashLine(0,0)(20,-20){4}
    \ArrowArc(0,-30)(15,40,140)
    \Text(20,15)[]{${\bar \lambda}^B_{{\dot\beta}} $}
    \Text(-25,-15)[]{${\bar \Psi}^C _{\dot\alpha}$}
    \Text(25,-15)[]{$A^A$}
    \end{picture}
\hspace{4.5cm}
-{\sqrt{2}\over 2}g f^{ABC}\epsilon_{{\dot \alpha}{\dot\beta}},
\end{eqnarray}
\vspace{0.3cm}
\begin{eqnarray}
\SetScale{0.8}
    \begin{picture}(80,20)(0,0)
        \Vertex(0,0){2}
    \ArrowLine(0,0)(0,20)
    \DashLine(0,0)(-20,-20){2}
    \DashLine(0,0)(20,-20){4}
    \ArrowArc(0,-30)(15,40,140)
    \Text(20,15)[]{${\bar \lambda}^B_{{\dot\alpha}} $}
    \Text(-25,-15)[]{$\Phi^C _\gamma$}
    \Text(29,-15)[]{$B^A,k$}
    \end{picture}
\hspace{2.7cm}
-{i\sqrt{2}\over 2}g d^{ABC}C^{\alpha\beta}\epsilon_{\beta\gamma}
\sigma^\mu_{\alpha{\dot \alpha}}k_\mu,
\end{eqnarray}
\vspace{0.3cm}
\begin{eqnarray}
\SetScale{0.8}
    \begin{picture}(50,0)(0,0)
    \Vertex(0,0){2}
    \DashLine(0,0)(20,20){2}
    \DashLine(0,0)(-20,20){4}
    \Gluon(-20,-20)(0,0){2}{4}
    \ArrowLine(0,0)(20,-20)
    \ArrowArc(0,30)(15,230,-50)
    \LongArrow(-25,-15)(-20,-8)
    \Text(-30,15)[]{$A^A$}
    \Text(30,15)[]{$\Psi^B_\alpha$}
    \Text(-30,-15)[]{$A_{\mu}^{C}$}
    \Text(30,-15)[]{${\bar \lambda}^D_{{\dot\gamma}}$}
    \end{picture}
\hspace{0.3cm} {\sqrt{2}\over 2}g^2\epsilon_{\alpha\beta}C^{\beta\gamma}
\sigma^{\mu}_{\gamma{\dot \gamma}}\times
(-3f^{ABL}d^{LCD}+f^{ACL}d^{LBD}\cr
+f^{BDL}d^{LAC}+f^{CDL}d^{LAD}),
\end{eqnarray}
\vspace{0.3cm}
\begin{eqnarray}
\SetScale{0.8}
    \begin{picture}(50,0)(0,0)
    \Vertex(0,0){2}
    \DashLine(0,0)(20,20){2}
    \DashLine(0,0)(-20,20){2}
    \ArrowLine(0,0)(-20,-20)
    \ArrowLine(0,0)(20,-20)
    \Text(-30,15)[]{$\zeta_{\alpha}^A$}
    \Text(30,15)[]{$\eta^B_\beta$}
    \Text(-30,-15)[]{${\bar \lambda}^C_{{\dot\alpha}}$}
    \Text(30,-15)[]{${\bar \lambda}^D_{{\dot\beta}}$}
    \end{picture}
\hspace{0.0cm} -{1\over 24}g^3 |C|^2\epsilon^{\alpha\beta}
\epsilon^{\dot{\alpha}\dot{\beta}}
\bigg\{\!\!\!\!\!\!&-&if^{ACM}d^{BDM}+if^{BCM}d^{ADM}\cr &-
&if^{ADM}d^{BCM}+if^{BDM}d^{ACM}\cr
&-&\;d^{ADM}d^{BCM}\,+\;d^{BCM}d^{ADM}\bigg\},
\end{eqnarray}
\vspace{.5cm}
where $A^a=(b,c,{\bar b},{\bar c})^a$, $\Psi^c_\alpha=(\zeta,\eta,\zeta,
\eta)^c_\alpha$ and $B^a=({\bar c},{\bar b})^a$,
$\Phi^c_{\alpha}=(\eta,\zeta)^c_\alpha$.
\section{One-loop Perturbative Corrections}
In this section, we will calculate the one-loop perturbative corrections to the 
undeformed ${\cal{N}}=1$ part and the $C$-deformed part of the 
theory separately. Due to the additional vertices more diagrams are to be 
considered comparing to the ordinary ${\cal{N}}=1$ supersymmetric $U(N)$. 
However, the tadpole diagrams from both undeformed and $C$-deformed sectors 
vanish identically due to the antisymmetry properties of $f^{abc}$ and 
$C^{\mu\nu}$. 
We will show that the quantum corrections of ${\cal 
N}=1$ part of the theory are not affected by $C$-deformation, whereas $C$ 
receives one-loop perturbative corrections. 
\subsection{${\cal N}=1$ Sector}
In this sector, the  standard field theory results for one-loop perturbative 
corrections to  ${\cal N}=1$ 
supersymmetric $U(N)$ gauge theory can be used
(for example see \cite{West}). Evaluating in particular 
the vertex function of the theory using $D$-dimensional regularization, the 
renormalization constant $Z_{1}$ is given by
\begin{eqnarray}
Z_{1}&=&Z_{A\bar{\lambda}\lambda}=1-\frac{Ng^{2}}{16\pi^{2}}\ 
\frac{2}{\epsilon}\;.
\end{eqnarray}
Further evaluating the self energy and the vacuum polarization tensor, 
the wave function renormalization constant 
$Z_{i}, i=2,3$ for $\lambda^{a}, \bar{\lambda}^{a}$ and $A_{\mu}^{a}$ can be 
determined and read  
\begin{eqnarray}
Z_2&=&Z^{\lambda{\bar \lambda}}
=1-{Ng^2\over 16\pi^2}\;{2\over \epsilon},\nonumber\\ 
Z_3&=&Z^{AA}=1+{3Ng^2\over 16\pi^2}\;{2\over \epsilon}.
\end{eqnarray}
Here $\epsilon=4-D$ is the regulator. Combining the above 
renormalization constants in the standard way, the gauge coupling  
renormalization constant $Z_{g}$  can be determined which is
\begin{eqnarray}\label{yyy}
Z_g&=&{Z_1\over Z_2\ Z_3^{1/2}}=1-{3\over 2}\;{Ng^2\over 16\pi^2}\;
{2\over \epsilon}.
\end{eqnarray}
Using now the relation $g_{0}=g\mu^{\epsilon/2}Z_{g}$ and the
fact that the bare coupling constant $g_{0}$ does not depend on the 
renormalization  scale 
$\mu$,  $\frac{d}{d\mu}g_{0}=0$, the $\beta$ function of the theory
can be obtained as 
\be\label{beta}
\beta(g(\mu))\equiv \mu\frac{\partial g(\mu)}{\partial\mu}=-{3Ng^3\over 16 
\pi^2}.
\ee
Solving this differential equation for the running coupling we arrive at
\begin{eqnarray}
g^{2}(\mu)=\frac{g^{2}(\mu_{0})}
{{3Ng^{2}(\mu_{0})\over 8\pi^2}\ln\frac{\mu}{\mu_{0}}+1},
\end{eqnarray}
where $\mu_{0}$ is a fixed energy.
\subsection{$C$-Deformed Sector}
The non(anti)commutativity parameter $C^{\mu\nu}$ appears in two different 
terms of the $C$-deformed part of the action. In the first term including the 
three fields interaction $A_{\mu}\bar{\lambda}\bar{\lambda}$, it appears as a rank 
two tensor contracted with $F_{\mu\nu}$ and in the second term containing a 
four $\bar{\lambda}$ interaction, it appears as a determinant. In the 
following subsection we will study the one-loop perturbative corrections 
corresponding  to both terms separately.  
\subsubsection{One-loop Correction to $C$; 
$A_{\mu}\bar{\lambda}\bar{\lambda}$ Three-Point Function}
In this section,  we will study in detail those one loop graphs 
including only one new vertex arising from the $C$-deformed part of the action. 
Starting with the two point function $\bar{\lambda}\bar{\lambda}$, the only 
possible diagram is the self-energy diagram from figure 1, with internal gluon 
and ghost fields.  Using the Feynman rules given in 
(\ref{FX1})-(\ref{FX3d}), one finds
\bea
\Gamma^{AB}_{{\dot \alpha}{\dot \beta}}\sim i g\; d^{ALC}f^{BLC}\;C^{\xi \mu}
\;\epsilon^{{\dot \alpha}{\dot \delta}}\;\sigma^{\rho}_{\gamma{\dot \delta}}\;
{\bar \sigma}_\mu^{{\dot\beta}\gamma}\;\int {d^4k\over (2\pi)^4}\;
{k_\rho(p-k)_{\xi} \over k^2(p-k)^2}
=0\;,
\eea
as expected. Here we have used the relation 
$C^{\xi\mu}\bar{\sigma}_{\mu}^{\ \dot{\beta}\gamma}\sigma^{\rho}_{\ 
\gamma\dot{\delta}}=-C^{\xi\rho}\delta^{\dot{\beta}}_{\ \dot{\delta}}$ and 
the antisymmetry property of $C^{\sigma\rho}$. Similarly one 
can also evaluate the two point function with 
ghost field in the loop. But, such a self energy diagram vanishes too, due to 
the same symmetry properties of $C^{\mu\nu}$ as above. Hence the two-point 
function 
$\bar{\lambda}\bar{\lambda}$ do not receive any  perturbative correction at 
one-loop order. Therefore the theory survives this first check of 
its renormalizability, at least at one-loop order. 
\par
The next step would be to consider the $C$-deformed  
$A\bar{\lambda}\bar{\lambda}$ three-point function of 
the theory. Evaluating
these three-point functions could indicate whether $C$ receives any one-loop 
correction or not.  All the relevant diagrams, containing only matter fields, 
are depicted in figure 2.
\par
Explicit calculation shows that the graphs (b,c,d),  figure 2, vanish 
identically taking into account the corresponding diagrams with crossed 
external legs. Equivalently, we can consider two different diagrams for each 
of the graphs
b,c, and d, and move within the loop first  in the clockwise direction 
and then in the counterclockwise direction.  After our convention for 
the Feynman rule
(\ref{FX4}), moving from $A_\mu$ to ${\bar \lambda}_{{\dot \alpha}}$
and then to ${\bar \lambda}_{{\dot \beta}}$ picks a plus sign and moving
in the other
direction, {\it i.e.} from  $A_\mu$ to ${\bar \lambda}_{{\dot \beta}}$
and then to ${\bar \lambda}_{{\dot \alpha}}$, a minus sign. Adding both 
contributions one can show that the diagrams (b,c), and (d) of figure 2 vanish. 
We are therefore left with diagram (a) from figure 2. To evaluate 
this graph we consider  three inequivalent situations depending on the position 
of the new $C$-deformed vertex in the graph. Let us label the vertices of the 
graph with (1,2,3) as shown in figure 3. Let us also indicate the position of 
the new vertex with an index ''c'' on the label of the vertices. We will find
therefore three different situations corresponding to $(1_c,2,3), (1,2_c,3)$ 
and  $(1,2,3_c)$.
\par
Using the Feynman rules (\ref{FX1})-(\ref{FX3d}) one can proceed to
evaluate these graphs. As a sample calculation let us compute the
graph corresponding to $(1_c,2,3)$
\bea
\Gamma_{({\rm a})}^{ABC,{\dot \alpha}{\dot\beta},\mu}&=&-2\times {1\over 3}
\times {3\over 2}g^3f^{ICJ}
d^{JAK}f^{KBI}C^{\rho\nu}\epsilon^{{\dot \alpha}{\dot \gamma}}
{\bar \sigma}^{\mu\; {\dot \delta}\gamma}{\sigma}^{\lambda}_{\gamma
{\dot \gamma}}
{\bar \sigma}^{\nu\; {\dot \beta}\delta}{\sigma}^{\kappa}_{\delta
{\dot \delta}}\cr
&&\cr &&
\int{d^4k\over (2\pi)^4}
\;\frac{k_\rho(k+p_1)_\lambda(k-p_2)_{\kappa}}{k^2(k+p_1)^2(k-p_2)^2}\;.
\eea
The factor 2 comes from two different directions  one can move in the
loop\footnote{Note that moving in 
two different directions within the loop in the diagram corresponding to 
$(1_{c},2,3)$, is equivalent to crossing the external legs of 
the diagram corresponding to $(1,2_c,3)$.} and $1/3$ is the symmetry factor. 
Using the 
conventions (\ref{convension}) and the fact that 
$C^{\rho\nu}
{\bar \sigma}^{\nu\; {\dot \beta}\delta}{\sigma}^{\kappa}_{\delta{\dot \delta}}
=-C^{\rho\kappa}\delta^{{\dot \beta}}_{{\dot \delta}}$, one finds
\begin{eqnarray}
\Gamma_{({\rm a})}^{ABC,{\dot \alpha}{\dot\beta},\mu}&=&
-{N\over 
2}g^3d^{ABC}
d_Af_Bf_C C^{\rho\kappa}\epsilon^{{\dot \alpha}{\dot \gamma}}
{\bar \sigma}^{\mu\; {\dot \beta}\gamma}{\sigma}^{\lambda}_{\gamma{\dot \gamma}}
(p_2)_{\kappa}\nonumber\\
&&\times\int{d^4k\over (2\pi)^4}
\;\frac{k_\rho(k+p_1)_\lambda}{k^2(k+p_1)^2(k-p_2)^2}\;.
\end{eqnarray}
The divergent part of the integral is ${1\over 4}{\delta^{\rho\lambda}\over
16\pi^2}\;{2\over \epsilon}$. Therefore we obtain 
\bea
\Gamma_{({\rm a})}^{ABC,{\dot \alpha}{\dot\beta},\mu}&=&{1\over 8}\;{Ng^3\over
16\pi^2}\; d^{ABC}
d_Af_Bf_C \;\epsilon^{{\dot \alpha}{\dot \beta}}\; C^{\mu\nu}(p_2)_\nu\;{2\over
\epsilon}+\mbox{finite terms}.
\eea
Adding the same contribution with $p_{1}\leftrightarrow p_{2}$ to this 
result, the final result for the divergent part of $(1_{c},2,3)$ reads
\bea
(1_c,2,3): &&\Gamma_{({\rm a})}^{ABC,{\dot \alpha}{\dot\beta},\mu}
={1\over 8}\;{Ng^3\over
16\pi^2}\; d^{ABC}
d_Af_Bf_C \;\epsilon^{{\dot \alpha}{\dot \beta}}\;
C^{\mu\nu}q_\nu\;{2\over \epsilon}\;,
\label{A1}
\eea
where $q=p_1+p_2$. Similarly the diagrams corresponding to two other situations 
$(1,2_c,3)$ and 
$(1,2,3_c)$ can be  evaluated explicitly. We find that
\bea
(1,2_c,3):&&\Gamma_{({\rm a})}^{ABC,{\dot \alpha}{\dot\beta},\mu}
={1\over 8}\;{Ng^3\over 16\pi^2}\; d^{ABC}
d_Bf_Af_C \;\epsilon^{{\dot \alpha}{\dot \beta}}\;
C^{\mu\nu}q_\nu\;{2\over \epsilon}\;,\cr &&\cr
(1,2,3_c):&&\Gamma_{({\rm a})}^{ABC,{\dot \alpha}{\dot\beta},\mu}
=-{1\over 2}\;{Ng^3\over 16\pi^2}\; d^{ABC}
d_Cf_Af_B \;\epsilon^{{\dot \alpha}{\dot \beta}}\;
C^{\mu\nu}q_\nu\;{2\over \epsilon}\;.
\label{A2}
\eea
\par
Let us now continue with the relevant ghost diagrams containing only one 
$C$-deformed vertex and contributing to the three point function 
$A_{\mu}\bar{\lambda}\bar{\lambda}$ (see figure 4).
As in the previous case each graph includes two different situations 
corresponding
to the position of the new $C$-deformed  vertex in the graph. Using the ghost 
Feynman rules one can evaluate these graphs as well. The results are
\bea
({\rm a})&&\left\{  \begin{array}{cc}
(1_c,2):\;\;\;\;\Gamma_{({\rm a})}^{ABC,{\dot \alpha}{\dot\beta},\mu}
={1\over 8}\;{Ng^3\over 16\pi^2}\; d^{ABC}
d_Af_Bf_C \;\epsilon^{{\dot \alpha}{\dot \beta}}\;
C^{\mu\nu}q_\nu\;{2\over \epsilon}\;,\\ \\
(1,2_c):\;\;\;\;\Gamma_{({\rm a})}^{ABC,{\dot \alpha}{\dot\beta},\mu}
={1\over 8}\;{Ng^3\over 16\pi^2}\; d^{ABC}
d_Bf_Af_C \;\epsilon^{{\dot \alpha}{\dot \beta}}\;
C^{\mu\nu}q_\nu\;{2\over \epsilon}\;,
\end{array}\right.\cr &&\cr &&\cr
({\rm b})&&\left\{  \begin{array}{cc}
(1_c,2):\;\;\;\;\Gamma_{({\rm b})}^{ABC,{\dot \alpha}{\dot\beta},\mu}
={1\over 4}\;{Ng^3\over 16\pi^2}\; d^{ABC}
d_Af_Bf_C \;\epsilon^{{\dot \alpha}{\dot \beta}}\;
C^{\mu\nu}q_\nu\;{2\over \epsilon}\;,\\ \\
(1,2_c):\;\;\;\;\Gamma_{({\rm b})}^{ABC,{\dot \alpha}{\dot\beta},\mu}
={1\over 4}\;{Ng^3\over 16\pi^2}\; d^{ABC}
d_Bf_Af_C \;\epsilon^{{\dot \alpha}{\dot \beta}}\;
C^{\mu\nu}q_\nu\;{2\over \epsilon}\;,
\end{array}\right.
\label{A3}
\eea
Having these results in hand, we are now ready to compute the counterterms and 
thereby
the corresponding  one-loop renormalization constant corresponding to the 
deformation parameter $C$. In 
fact looking at
the action we see that the $A_\mu{\bar\lambda}{\bar\lambda}$ term has 
different $U(1)$ and $SU(N)$ 
components 
\bea
&& i\sqrt{{2\over N}}g\epsilon^{{\dot \alpha}{\dot \beta}}
C^{\mu\nu}\partial_\mu A_\nu^0{\bar \lambda}_{{\dot \alpha}}^0
{\bar \lambda}_{{\dot \beta}}^0+
i\sqrt{{2\over N}}g\epsilon^{{\dot \alpha}{\dot \beta}}\delta^{ab}
C^{\mu\nu}\partial_\mu A_\nu^0{\bar \lambda}_{{\dot \alpha}}^a
{\bar \lambda}_{{\dot \beta}}^b+igd^{abc}
\epsilon^{{\dot \alpha}{\dot \beta}}
C^{\mu\nu}\partial_\mu A_\nu^c{\bar \lambda}_{{\dot \alpha}}^a
{\bar \lambda}_{{\dot \beta}}^b\cr &&\cr
&&+i\sqrt{{2\over N}}g\epsilon^{{\dot \alpha}{\dot \beta}}\delta^{ca}
C^{\mu\nu}\partial_\mu A_\nu^c{\bar \lambda}_{{\dot \alpha}}^a
{\bar \lambda}_{{\dot \beta}}^0+
i\sqrt{{2\over N}}g\epsilon^{{\dot \alpha}{\dot \beta}}\delta^{cb}
C^{\mu\nu}\partial_\mu A_\nu^c{\bar \lambda}_{{\dot \alpha}}^0
{\bar \lambda}_{{\dot \beta}}^b\;.
\label{B1}
\eea
Adding the different contributions (\ref{A1}), (\ref{A2}) and (\ref{A3}) from 
the diagrams of figure 3 and 4 
corresponding to different $U(1)$ and $SU(N)$ couplings we arrive at the 
following counterterms for the five possible combinations of 
$A_{\mu}^{C}-\bar{\lambda}^{A}-\bar{\lambda}^{B}$ in (\ref{B1}). Whereas the 
first term including $(U(1))^{3}$ coupling receives no quantum correction, 
all the other counterterms can be given by
 \begin{eqnarray}\label{nn}
A^{0}\bar{\lambda}^{a}\bar{\lambda}^{b}:&&-\frac{g^{2}N}{16\pi^2}\ 
\frac{2}{\epsilon},\nonumber\\
A^{c}\bar{\lambda}^{a}\bar{\lambda}^{0}:&&+\frac{g^{2}N}{16\pi^{2}}\ 
\frac{2}{\epsilon},\nonumber\\
A^{c}\bar{\lambda}^{0}\bar{\lambda}^{b}:&&+\frac{g^{2}N}{16\pi^{2}}\ 
\frac{2}{\epsilon},\nonumber\\
A^{c}\bar{\lambda}^{a}\bar{\lambda}^{b}:&&+\frac{1}{2}\ 
\frac{g^{2}N}{16\pi^{2}}\ 
\frac{2}{\epsilon}. 
\end{eqnarray}   
Adding now the counterterm action to the original one and comparing the 
resulting 
expression with the bare action the value of one-loop perturbative correction 
to $C$ can be determined. 
Using the relation $C^{\mu\nu}_{0}\equiv Z_{C}C_{\mu\nu}$ between the bare  
parameter $C_{0}^{\mu\nu}$ and the renormalized parameter $C^{\mu\nu}$, one  
finds immediately 
from the first uncorrected term including $(U(1))^{3}$ coupling that 
$Z_{C}=Z_{g}^{-1}$.  Using further the 
results from the one-loop computations of the undeformed ${\cal{N}}=1$ 
sector  and the $C$-deformed sector (\ref{nn}), one finds the 
corresponding corrections to the other terms of the action including two or 
three  $SU(N)$ couplings  
\bea
A^0{\bar \lambda}^a{\bar \lambda}^b: &\;\;\;\;\;&\left
(1-{g^2N\over 16\pi^2}\;{2\over \epsilon}\right)= Z_2\;,\cr &&\cr
A^c{\bar \lambda}^a{\bar \lambda}^0: &\;\;\;\;\;&\left
(1+{g^2N\over 16\pi^2}\;{2\over \epsilon}\right)= (Z_2Z_3)^{1/2}\;,
\cr &&\cr
A^c{\bar \lambda}^0{\bar \lambda}^b: &\;\;\;\;\;&\left
(1+{g^2N\over 16\pi^2}\;{2\over \epsilon}\right)= (Z_2Z_3)^{1/2}\;,
\cr &&\cr
A^c{\bar \lambda}^a{\bar \lambda}^b: &\;\;\;\;\;&\left
(1+{1\over 2}\;{g^2N\over 16\pi^2}\;
{2\over \epsilon}\right)= (Z_2^2Z_3)^{1/2}\;.
\eea
Comparing again each term with the bare action, all these contributions 
cancel and we are 
left with the conclusion that $Z_{C}=Z_{g}^{-1}$ for all $U(1)$ and $SU(N)$ 
couplings. The gauge coupling renormalization constant  $Z_{g}$ is 
determined from the undeformed ${\cal{N}}=1$ 
sector of the theory (\ref{yyy}). Using this result, the one-loop quantum 
correction to the non(anti)commutativity parameter $C$ reads
\be
C_{0}^{\mu\nu}=\left(1+{3\over 2}\;{g^2N\over 16\pi^2}\;
{2\over \epsilon}\right)C^{\mu\nu}\equiv Z_C C^{\mu\nu}\;.
\label{bbb}
\ee
\subsubsection{One-loop Correction to $C^2$; ${\bar \lambda}$ 
Four-Point Function}
The non-trivial $\bar{\lambda}$ four-point functions
containing only matter fields and at most two $C$-deformed vertices are 
depicted in figure 5. They lead to one-loop perturbative corrections to 
the  coupling $|C|^2$.
Explicit calculations show that the graphs (a,b,c) are zero 
when their corresponding crossed graphs are taken  
into account. The situation is very similar to that in three-point
functions mentioned in the previous section. Therefore as far as the matter
fields are concerned the corrections come from graph (d) in figure 5.
More precisely, there are two different situations for this graph. They are 
given in figure 6. To evaluate the
graphs in figure 6 we have to consider different inequivalent situations for 
each
graph depending on the position of the new C-dependent vertex in the graph. We 
obtain
\bea
(1_c,2,3_c,4):\hspace{0.7cm}
\Gamma_{(6a)}^{ABCD,\dot{\alpha}\dot{\beta}\dot{\gamma}\dot{\delta}}
&=&0,\hspace{1cm}
(1,2_c,3,4_c):\hspace{.5cm}
\Gamma_{(6a)}^{ABCD,\dot{\alpha}\dot{\beta}\dot{\gamma}\dot{\delta}}=0,
\cr &&\cr
(1_c,2,3_c,4):\hspace{.5cm}
\Gamma_{(6b)}^{ABCD,\dot{\alpha}\dot{\beta}\dot{\gamma}\dot{\delta}}
&=&0,\hspace{1cm}
(1,2_c,3,4_c):\hspace{0.5cm}
\Gamma_{(6b)}^{ABCD,\dot{\alpha}\dot{\beta}\dot{\gamma}\dot{\delta}}
=0\;.
\eea
The remaining nonzero graphs are all equal. For instance, using the Feynman 
rules from (\ref{FX1})-(\ref{FX3d}), the Feynman integral 
corresponding 
to the $(1_c, 2_{c},3,4)$ situation is given by
\bea
(1_c,2_c,3,4):\hspace{0.5cm}\Gamma_{(a)}^{ABCD,\dot{\alpha}\dot{\beta}
\dot{\gamma}\dot{\delta}}&=&{9\over 4}g^4\;
d^{IAJ}d^{JBK}f^{KCL}f^{LDI}\;
C^{\lambda\mu}{C_{\mu}}^{\rho}(\bar{\sigma}^\nu\sigma^\chi\epsilon)^
{\dot{\delta}\dot{\alpha}}(\bar{\sigma}_\nu\sigma^\kappa\epsilon)^
{\dot{\gamma}\dot{\beta}}\cr &&\cr
&\times&\int {d^4k\over (2\pi)^4}
{k_\lambda k_\rho (k+p_2)_\kappa
(k-p_1)_\chi \over k^2(k-p_1)^2(k+p_2)^2(k-p_1-p_4)^2}\;.
\eea
Taking into account all the
symmetry factors and summing over all different configurations, the divergent 
part of the above expression is given by
\be
\Gamma^{ABCD,\dot{\alpha}\dot{\beta}\dot{\gamma}\dot{\delta}}=
-{1\over 16\pi^2}({8\times 4\times 2\over 4!})({9\over 8})C^2g^4
\epsilon^{\dot{\alpha}\dot{\beta}}\epsilon^{\dot{\gamma}\dot{\delta}}
d^{IAJ}d^{JBK}f^{KCL}f^{LDI}\;{2\over \epsilon}\;.
\ee
In particular, for the situation where two of ${\bar \lambda}$'s carry $U(1)$
index and the others $SU(N)$ index the above expression reads
\be
\Gamma^{ab00,\dot{\alpha}\dot{\beta}\dot{\gamma}\dot{\delta}}=
{3\over 8\pi^2}\  C^2g^4\epsilon^{\dot{\alpha}\dot{\beta}}
\epsilon^{\dot{\gamma}\dot{\delta}}\delta^{ab}{2\over \epsilon}\;.
\label{ooo}
\ee
This is the one-loop correction to 
$\frac{1}{4N}g^2|C|^2\epsilon^{\dot{\alpha}\dot{\beta}}
\epsilon^{\dot{\gamma}\dot{\delta}}\delta^{ab}{\bar 
\lambda}^0{\bar 
\lambda}^0
{\bar \lambda}^a{\bar \lambda}^b$ term in the action arising from the matter 
fields in $(U(1))^{2}\times (SU(N))^{2}$ combination. 
\par
Now what concerns the ghost field contributions to the $\bar{\lambda}$ 
four-point 
function, explicit calculations show that all diagrams depicted in figure 7, 
containing two $C$-deformed vertices vanish due to the antisymmetry property 
of $C^{\mu\nu}$ two-form, and therefore (\ref{ooo})
is the only one-loop correction to the ${\bar \lambda}$
four-point function for the $(U(1))^{2}\times (SU(N))^{2}$ coupling. To 
determine the  corresponding correction 
to the action (counterterm), a factor $-1/24$ must be multiplied with the 
above 
$\bar{\lambda}$ four-point correction (\ref{ooo}). This factor arises from 24 
different 
contractions
which are to be performed to obtain the four ${\bar \lambda}$ interaction
vertex. Equivalently, the above correction (\ref{ooo}) can be compared 
with the four $\bar{\lambda}$ vertex and leads to the counterterm 
to  ${\bar \lambda}^4$ part of the action.
\par
Next adding the part of the original and the counterterm action containing 
$\bar{\lambda}^{0}\bar{\lambda}^{0}\bar{\lambda}^{a}\bar{\lambda}^{b}$ and 
comparing the resulting expression with the bare action containing the same 
$\bar{\lambda}$ combination, we conclude that $Z_{C^{2}}=Z_{g}^{-2}$. Using 
further 
the one-loop results from previous section $Z_{C}=Z_{g}^{-1}$, we arrive at the 
conclusion that
\begin{eqnarray}
Z_{C^{2}}=Z_{C}^{2}.
\end{eqnarray}
This results was also suggested in \cite{Lunin:2003bm}, where it was supposed 
to be correct in all order of perturbation theory due to  some consistency 
arguments. Using the value of $Z_{g}$ from (\ref{yyy}) we arrive at
\be
C_{0}^2=\left(1+3\;{g^2N\over 16\pi^2}\;
{2\over \epsilon}\right)\;C^2\equiv Z_{C^2}C^2\;.
\ee
\section{Running $C^{\mu\nu}$ and $C^{2}$}
As next it is interesting to study the running of the 
non(anti)commutativity parameter $C^{\mu\nu}$ 
and the related $|C|^{2}$ with the renormalization  scale $\mu$. 
Let us consider the 
bare three-point function $\Gamma^{0}_{A\bar{\lambda}\bar{\lambda}}$ consisting 
of one gauge and two $\bar{\lambda}$ fermion fields. The Callan-Symanzik 
differential equation
for the corresponding renormalized three-point function, 
$\Gamma^{R}_{A\bar{\lambda}\bar{\lambda}}$, can be given
using the fact that the bare three-point function, 
$\Gamma^{0}_{A\bar{\lambda}\bar{\lambda}}$, is
independent of the renormalization scale $\mu$. In other words, one has
\begin{eqnarray}
\Gamma^{0}_{A\bar{\lambda}\bar{\lambda}}=Z_{3}^{-1/2}Z_{2}^{-1}
\ \Gamma^{R}(p_{i}, g(\mu),
C^{\mu\nu}(\mu); \mu), \hspace{1cm}
\mu\frac{d}{d\mu}\Gamma^{0}_{A\bar{\lambda}\bar{\lambda}}=0\;,
\end{eqnarray}
which leads to the following differential equation for the renormalized 
three-point function $\Gamma_{A\bar{\lambda}\bar{\lambda}}^{R}$
\begin{eqnarray}
\left(\mu\frac{\partial}{\partial \mu}\ln Z_{3}^{-1/2}+2
\mu\frac{\partial}{\partial \mu}\ln Z_{2}^{-1/2}+\mu\frac{\partial}
{\partial \mu}+\beta(g)\frac{\partial}{\partial g}+C^{\mu\nu}\ \gamma_{C}(g)
\frac{\partial}{\partial 
C^{\mu\nu}}\right)\Gamma^{R}_{A\bar{\lambda}\bar{\lambda}}=0\;.
\end{eqnarray}
Here $\beta(g)$ is the ordinary $\beta$-function of the theory (\ref{beta}) 
and $\gamma_{C}$ is defined by
\begin{eqnarray}
C^{\mu\nu}\gamma_{C}(g)\equiv \mu\frac{\partial}{\partial \mu}
C^{\mu\nu}(\mu)\;.
\end{eqnarray}
To see how the non(anti)commutativity parameter $C^{\mu\nu}$ runs with 
the renormalization scale $\mu$, it is useful to find a general expression for 
$\gamma_{C}$ 
in terms
of $Z_{C}$. Using (\ref{bbb}) we arrive at
\begin{eqnarray}
C^{\mu\nu}\gamma_{C}(g)=\mu\frac{\partial}
{\partial\mu}\left(Z_{C}^{-1}\right)C^{\mu\nu}_{0}=C^{\mu\nu}
Z_{C}\ \mu\frac{\partial}{\partial \mu}Z_{C}^{-1}\;,
\end{eqnarray}
which means that
\be
\gamma_{C}=-\mu\frac{\partial}{\partial \mu}\ln Z_{C}=-\beta(g){\partial
\over \partial g}\ln Z_{C}\;.
\ee
Using our previous results for $Z_{C}$, we obtain 
\be
\gamma_{C}=+\frac{3Ng^{2}}{16\pi^{2}}\;.
\ee
It is now easy to solve this equation and study the behavior of
non\-(anti)\-com\-mu\-ta\-ti\-vi\-ty parameter in terms of the scale  $\mu$.
Indeed, using the fact that the gauge coupling runs as
$g^{2}(\mu)=\frac{8\pi^{2}}{3N\ln(\mu/\Lambda_{U(N)})}$, with
$\Lambda_{U(N)}=\mu_0\ e^{-{8\pi^2\over 3N g^2(\mu_0)}}$ \footnote{
We note that $\Lambda_{U(N)}$ is RG invariant, {\it i.e.}
$\Lambda_{U(N)}=\mu_0 e^{-{8\pi^2\over 3N g^2(\mu_0)}}=
\mu e^{-{8\pi^2\over 3N g^2(\mu)}}$.}, we get
\begin{eqnarray}
\ln\frac{C^{\mu\nu}(\mu)}{C^{\mu\nu}(\mu_{0})}=\frac{1}{2}
\int\limits_{\mu_{0}}^{\mu}
\frac{d\mu}{\mu\ln\frac{\mu}{\Lambda_{U(N)}}}=\ln
\left(\frac{\ln(\mu/\Lambda_{U(N)})}{\ln(\mu_{0}/
\Lambda_{U(N)})}\right)^{1/2}\;,
\end{eqnarray}
which leads to
\begin{eqnarray}\label{square}
C^{\mu\nu}(\mu)g(\mu)=C^{\mu\nu}(\mu_{0})g(\mu_{0})\;.
\end{eqnarray}
We arrive therefore at
\begin{eqnarray}
C^{2}(\mu)g^{2}(\mu)=C^{2}(\mu_{0})g^{2}(\mu_{0})={\rm const}\;.
\label{xxx}
\end{eqnarray}
Alternatively, the same relation can be obtained by looking for an appropriate 
Callan-Symanzik for the renormalized $\bar{\lambda}$ four-point function of the 
theory and 
solving the equation
for running $C^{2}(\mu)$ using the fact that $Z_{C^{2}}=(Z_{C})^{2}$. This 
leads to 
$\gamma_{C^{2}}=2\gamma_{C}$ and consequently the same result 
from (\ref{xxx}) can be obtained.  
\par
An immediate consequence of the equation (\ref{xxx}) is that another RG 
invariant, $\Lambda_{C}$,  can be defined which is related to $C^{2}$ at a 
fixed energy  and can be expressed as a 
function of the well-known RG invariant $\Lambda_{U(N)}$ 
\be
\Lambda_{C}=\frac{1}{C^{2}(\mu_{0})}\;\ln\left({\mu_0\over
\Lambda_{U(N)}}\right)
=\frac{1}{C^{2}(\mu)}\;\ln\left({\mu\over \Lambda_{U(N)}}\right)\;.
\ee
One can show that $\Lambda_{C}$ is always positive for 
$\mu_{0}\gg\Lambda_{U(N)}$\footnote{We do not consider the case where 
$\mu_{0}=\Lambda_{U(N)}$, because in this case  
$g(\mu_{0}=\Lambda_{U(N)})\to \infty$, and this  would break our perturbative 
considerations. Further the case $\mu_{0}<\Lambda_{U(N)}$ would end up with an 
imaginary gauge coupling and this breaks the unitarity of the theory.}.  
This means that $C^{2}$ grows linearly with $\ln(\mu/\Lambda_{U(N)})$ 
with the slope
$\Lambda_{C}^{-1}$. Using further the fact that the running 
coupling  $g^{2}(\mu)=
\frac{8\pi^{2}}{3N\ln(\mu/\Lambda_{U(N)})}$, we obtain
\begin{eqnarray}
C^{2}(\mu)g^{2}(\mu)=\frac{8\pi^2}{3N \Lambda_{C}}
\equiv\eta^{2}={\rm const.}
\end{eqnarray}
Due to the positivity of 
$\Lambda_{U(N)}$, the constant $\eta$ is always positive. 
In fact $\eta$ is the value of the gauge
coupling at the energy where the running  coupling constant and 
non(anti)commutativity parameter have the same value. If we plot $C^2$ and 
$g^2$ in one and the same diagram as a function of $\ln\mu$, both curves 
intersect at one point, say $\mu_c$, and we obtain $g(\mu_c)=\eta$.

\section{Conclusion}

We have studied perturbative corrections to pure ${\cal N}={1\over 2}$ 
supersymmetric $U(N)$ gauge theory at one loop order. We have used the fact 
that the corresponding action to this theory can be 
separated into two parts: the first part  preserves standard 
${\cal N}=1$ supersymmetry and 
the second part is the $C$-deformed part and 
breaks the supersymmetry to  ${\cal N}={1\over 2}$. We have shown that 
${\cal N}=1$ part does not receive any $C$-dependent corrections and
can therefore be treated as the standard ${\cal N}=1$ supersymmetric $U(N)$ 
gauge theory. Explicit one-loop calculation of $A\bar{\lambda}\bar{\lambda}$ 
three-point function  and $\bar{\lambda}$ four-point function show however 
that the  non(anti)commutativity parameter $C^{\mu\nu}$ and $C^{2}$ receive  
one-loop corrections.  As a result we have $Z_{C}=Z_{g}^{-1}$ with 
$Z_{g}$ the standard gauge coupling renormalization constant and consequently 
\be
Z_C^2=(Z_C)^2=1+3{Ng^2\over 16\pi}\;{2\over \epsilon}\;.
\ee
Using this correction we found the running of the
non(anti)commutativity parameter as a function of the renormalization
scale $\mu$
\be
C^2(\mu)={1\over \Lambda_C}\ln({\mu\over\mu_0})+C^2(\mu_0)\;.
\ee
where $\Lambda_{C}$ is the new RG invariant scale.  

Finally we note that since the ${\cal N}=1$ sector of the theory remains
unaffected by the $C$-deformation, the standard anomaly of the theory
would be the same as before. Nevertheless one would expect to get a $C$-dependent
corrections for those anomalies which whole supersymmetry of the theory
(${\cal N}=1/2$) is involved. In particular the Konishi anomaly can also
be studied along \cite{Ardalan:2003ev} (see also \cite{Kawai:2003yf})\cite{Sadooghi}. 
 
\vspace*{.4cm}
\subsubsection*{Acknowledgements}
We would like to thank Farhad Ardalan, Hessam Arfaei and Amir Mosaffa
for useful comments. Special thanks to Shahrokh Parvizi for useful
discussions.
\newpage
\section{Figures}
\begin{eqnarray}
\SetScale{0.8}
\begin{picture}(0,0)(0,0)
\Line(-75,0)(75,0)
\GlueArc(0,0)(25,0,180){2}{10}
\Vertex(-25,0){2}\Vertex(25,0){2}
\Text(-57,-10)[]{${\bar \lambda}_{{\dot \alpha}}^A$}
\Text(60,-10)[]{${\bar \lambda}_{{\dot \beta}}^B$}
\Text(0,-30)[]{Figure 1}\nonumber
\end{picture}
\end{eqnarray}
\vspace{1cm}
\begin{eqnarray}
\SetScale{0.8}
\begin{picture}(0,45)(0,0)
\Line(-200,50)(-100,0)
\Line(-200,-50)(-100,0)
\Gluon(-150,25)(-150,-25){2}{5}
\Gluon(-100,0)(-50,0){2}{5}
\Vertex(-150,25){2}\Vertex(-150,-25){2}\Vertex(-100,0){2}
\Text(-160,25)[]{${\bar \lambda}_{{\dot \alpha}}^A$}
\Text(-160,-25)[]{${\bar \lambda}_{{\dot \beta}}^B$}
\Text(-50,-15)[]{$A_\mu^C$}
\Text(-100,-75)[]{(a)}
\Line(50,50)(100,25)
\Line(50,-50)(100,-25)
\Gluon(100,25)(150,0){2}{5}
\Gluon(100,-25)(150,0){2}{5}
\Line(100,25)(100,-25)
\Gluon(150,0)(200,0){2}{5}
\Vertex(100,25){2}\Vertex(100,-25){2}\Vertex(150,0){2}
\Text(40,25)[]{${\bar \lambda}_{{\dot \alpha}}^A$}
\Text(40,-25)[]{${\bar \lambda}_{{\dot \beta}}^B$}
\Text(150,-15)[]{$A_\mu^C$}
\Text(100,-75)[]{(b)}
\GlueArc(-125,-150)(25,0,360){2}{20}
\Line(-200,-125)(-150,-150)
\Line(-200,-175)(-150,-150)
\Vertex(-150,-150){2}
\Gluon(-100,-150)(-50,-150){2}{5}
\Vertex(-100,-150){2}
\Text(-150,-95)[]{${\bar \lambda}_{{\dot \alpha}}^A$}
\Text(-150,-150)[]{${\bar \lambda}_{{\dot \beta}}^B$}
\Text(-50,-135)[]{$A_\mu^C$}
\Text(-100,-190)[]{(c)}
\Gluon(50,-125)(100,-150){2}{7}
\Line(50,-175)(100,-150)
\Line(100,-150)(200,-150)
\GlueArc(125,-150)(25,0,180){2}{10}
\Vertex(100,-150){2} \Vertex(150,-150){2}
\Text(40,-150)[]{${\bar \lambda}_{{\dot \alpha}}^A$}
\Text(40,-90)[]{$A_\mu^C$}
\Text(160,-135)[]{${\bar \lambda}_{{\dot \beta}}^B$}
\Text(100,-190)[]{(d)}
\Text(0,-230)[]{Figure 2}
\nonumber
\end{picture}
\end{eqnarray}
\vspace{8cm}
\begin{eqnarray}
\SetScale{0.8}
\begin{picture}(-230,45)(0,0)
\Line(-200,50)(-100,0)
\Line(-200,-50)(-100,0)
\Gluon(-150,25)(-150,-25){2}{5}
\Gluon(-100,0)(-50,0){2}{5}
\Vertex(-150,25){2}\Vertex(-150,-25){2}\Vertex(-100,0){2}
\Text(-160,25)[]{${\bar \lambda}_{{\dot \alpha}}^A$}
\Text(-160,-25)[]{${\bar \lambda}_{{\dot \beta}}^B$}
\Text(-50,-12)[]{$A_\mu^C$}
\Text(-80,8)[]{3}
\Text(-120,28)[]{1}
\Text(-120,-28)[]{2}
\Text(-100,-75)[]{Figure 3}
\nonumber
\end{picture}
\end{eqnarray}
\vspace{1.5cm}

\newpage

\begin{eqnarray}
\SetScale{0.8}
\begin{picture}(0,45)(0,0)
\Line(-200,50)(-150,25)
\DashLine(-150,25)(-100,0){2}
\Line(-200,-50)(-150,-25)
\DashLine(-150,-25)(-100,0){2}
\DashLine(-150,25)(-150,-25){4}
\Gluon(-100,0)(-50,0){2}{5}
\Vertex(-150,25){2}\Vertex(-150,-25){2}\Vertex(-100,0){2}
\Text(-160,25)[]{${\bar \lambda}_{{\dot \alpha}}^A$}
\Text(-160,-25)[]{${\bar \lambda}_{{\dot \beta}}^B$}
\Text(-50,-15)[]{$A_\mu^C$}
\Text(-100,-50)[]{(a)}
\Text(-125,-15)[]{2}\Text(-125,15)[]{1}
\Line(50,50)(100,25)
\Line(50,-50)(100,-25)
\DashLine(100,25)(150,0){4}
\DashLine(100,-25)(150,0){4}
\DashLine(100,25)(100,-25){2}
\Gluon(150,0)(200,0){2}{5}
\Vertex(100,25){2}\Vertex(100,-25){2}\Vertex(150,0){2}
\Text(40,25)[]{${\bar \lambda}_{{\dot \alpha}}^A$}
\Text(40,-25)[]{${\bar \lambda}_{{\dot \beta}}^B$}
\Text(150,-15)[]{$A_\mu^C$}
\Text(100,-50)[]{(b)}
\Text(75,-15)[]{2}\Text(75,15)[]{1}
\Text(0,-80)[]{Figure 4}
\nonumber
\end{picture}
\end{eqnarray}
\vspace{2.5cm}

\begin{eqnarray}
\SetScale{0.8}
\begin{picture}(0,55)(0,0)
\Line(-200,50)(-50,-25)
\Line(-200,-50)(-50,25)
\Gluon(-150,25)(-150,-25){2}{5}
\Vertex(-150,25){2}\Vertex(-150,-25){2}\Vertex(-100,0){2}
\Text(-160,25)[]{${\bar \lambda}_{{\dot \alpha}}^A$}
\Text(-160,-25)[]{${\bar \lambda}_{{\dot \beta}}^B$}
\Text(-50,25)[]{${\bar \lambda}_{{\dot \gamma}}^C$}
\Text(-50,-25)[]{${\bar \lambda}_{{\dot \delta}}^D$}
\Text(-100,-75)[]{(a)}
\Line(50,50)(100,25)
\Line(50,-50)(100,-25)
\Gluon(100,25)(150,0){2}{5}
\Gluon(100,-25)(150,0){2}{5}
\Line(100,25)(100,-25)
\Line(150,0)(200,25)
\Line(150,0)(200,-25)
\Vertex(100,25){2}\Vertex(100,-25){2}\Vertex(150,0){2}
\Text(40,25)[]{${\bar \lambda}_{{\dot \alpha}}^A$}
\Text(40,-25)[]{${\bar \lambda}_{{\dot \beta}}^B$}
\Text(150,25)[]{${\bar \lambda}_{{\dot \gamma}}^C$}
\Text(150,-25)[]{${\bar \lambda}_{{\dot \delta}}^D$}
\Text(100,-75)[]{(b)}
\GlueArc(-125,-150)(25,0,360){2}{10}
\Line(-200,-125)(-150,-150)
\Line(-200,-175)(-150,-150)
\Vertex(-150,-150){2}
\Line(-100,-150)(-50,-125)
\Line(-100,-150)(-50,-175)
\Vertex(-100,-150){2}
\Text(-150,-95)[]{${\bar \lambda}_{{\dot \alpha}}^A$}
\Text(-150,-150)[]{${\bar \lambda}_{{\dot \beta}}^B$}
\Text(-50,-95)[]{${\bar \lambda}_{{\dot \gamma}}^C$}
\Text(-50,-150)[]{${\bar \lambda}_{{\dot \delta}}^D$}
\Text(-100,-190)[]{(c)}
\Line(50,-125)(200,-125)
\Line(50,-175)(200,-175)
\Gluon(100,-125)(100,-175){2}{5}
\Gluon(150,-125)(150,-175){2}{5}
\Vertex(100,-125){2} \Vertex(100,-175){2}
\Vertex(150,-125){2}\Vertex(150,-175){2}
\Text(40,-90)[]{${\bar \lambda}_{{\dot \alpha}}^A$}
\Text(40,-150)[]{${\bar \lambda}_{{\dot \beta}}^B$}
\Text(160,-90)[]{${\bar \lambda}_{{\dot \gamma}}^C$}
\Text(160,-150)[]{${\bar \lambda}_{{\dot \delta}}^D$}
\Text(100,-190)[]{(d)}
\Text(0,-220)[]{Figure 5}\nonumber
\end{picture}
\end{eqnarray}
\vspace{7cm}

\begin{eqnarray}
\SetScale{0.8}
\begin{picture}(400,0)(0,-100)
\Line(50,-125)(200,-125)
\Line(50,-175)(200,-175)
\Gluon(100,-125)(100,-175){2}{5}
\Gluon(150,-125)(150,-175){2}{5}
\Vertex(100,-125){2} \Vertex(100,-175){2}
\Vertex(150,-125){2}\Vertex(150,-175){2}
\Text(80,-90)[]{1}\Text(120,-90)[]{4}
\Text(80,-150)[]{2}\Text(120,-150)[]{3}
\Text(40,-90)[]{${\bar \lambda}_{{\dot \alpha}}^A$}
\Text(40,-150)[]{${\bar \lambda}_{{\dot \beta}}^B$}
\Text(160,-90)[]{${\bar \lambda}_{{\dot \delta}}^D$}
\Text(160,-150)[]{${\bar \lambda}_{{\dot \gamma}}^C$}
\Text(100,-190)[]{(a)}
\Line(300,-125)(350,-125) \Line(400,-125)(450,-125)
\Line(300,-175)(350,-175) \Line(400,-175)(450,-175)
\Gluon(350,-125)(400,-125){2}{5}
\Gluon(350,-175)(400,-175){2}{5}
\Line(350,-125)(350,-175)
\Line(400,-125)(400,-175)
\Vertex(350,-125){2} \Vertex(400,-125){2}
\Vertex(350,-175){2}\Vertex(400,-175){2}
\Text(280,-90)[]{1}\Text(320,-90)[]{4}
\Text(280,-150)[]{2}\Text(320,-150)[]{3}
\Text(240,-90)[]{${\bar \lambda}_{{\dot \alpha}}^A$}
\Text(360,-90)[]{${\bar \lambda}_{{\dot \delta}}^D$}
\Text(240,-150)[]{${\bar \lambda}_{{\dot \beta}}^B$}
\Text(360,-150)[]{${\bar \lambda}_{{\dot \gamma}}^C$}
\Text(300,-190)[]{(b)}
\Text(200,-220)[]{Figure 6}
\nonumber
\end{picture}
\end{eqnarray}
\vspace{4cm}

\newpage
\begin{eqnarray}
\SetScale{0.8}
\begin{picture}(0,55)(0,0)
\Line(-200,50)(-150,25)\Line(-100,0)(-50,-25)
\Line(-200,-50)(-150,-25)\Line(-100,0)(-50,25)
\DashLine(-150,25)(-150,-25){2}
\DashLine(-150,25)(-100,0){4}
\DashLine(-150,-25)(-100,0){4}
\Vertex(-150,25){2}\Vertex(-150,-25){2}\Vertex(-100,0){2}
\Text(-160,25)[]{${\bar \lambda}_{{\dot \alpha}}^A$}
\Text(-160,-25)[]{${\bar \lambda}_{{\dot \beta}}^B$}
\Text(-50,25)[]{${\bar \lambda}_{{\dot \gamma}}^C$}
\Text(-50,-25)[]{${\bar \lambda}_{{\dot \delta}}^D$}
\Text(-100,-75)[]{(a)}
\Line(50,50)(100,25)
\Line(50,-50)(100,-25)
\DashLine(100,25)(150,0){2}
\DashLine(100,-25)(150,0){2}
\DashLine(100,25)(100,-25){4}
\Line(150,0)(200,25)
\Line(150,0)(200,-25)
\Vertex(100,25){2}\Vertex(100,-25){2}\Vertex(150,0){2}
\Text(40,25)[]{${\bar \lambda}_{{\dot \alpha}}^A$}
\Text(40,-25)[]{${\bar \lambda}_{{\dot \beta}}^B$}
\Text(150,25)[]{${\bar \lambda}_{{\dot \gamma}}^C$}
\Text(150,-25)[]{${\bar \lambda}_{{\dot \delta}}^D$}
\Text(100,-75)[]{(b)}
\nonumber
\end{picture}
\end{eqnarray}
\vspace{3cm}
\begin{eqnarray}
\SetScale{0.8}
\begin{picture}(-205,70)(0,0)
\Line(-200,50)(-150,50) \Line(-100,50)(-50,50)
\Line(-200,0)(-150,0) \Line(-100,0)(-50,0)
\DashLine(-150,50)(-150,0){2}
\DashLine(-100,50)(-100,0){2}
\DashLine(-150,50)(-100,50){4}
\DashLine(-150,0)(-100,0){4}
\Vertex(-150,50){2} \Vertex(-100,50){2}
\Vertex(-150,0){2}\Vertex(-100,0){2}
\Text(-170,50)[]{${\bar \lambda}_{{\dot \alpha}}^A$}
\Text(-30,50)[]{${\bar \lambda}_{{\dot \beta}}^B$}
\Text(-170,-10)[]{${\bar \lambda}_{{\dot \gamma}}^C$}
\Text(-30,-10)[]{${\bar \lambda}_{{\dot \delta}}^D$}
\Text(-100,-40)[]{(c)}
\Text(-100,-70)[]{Figure 7}\nonumber
\nonumber
\end{picture}
\end{eqnarray}

\end{document}